\documentclass[a4paper,11pt]{article}
\pdfoutput=1

\usepackage[normalem]{ulem}
\usepackage{jcappub} 
\usepackage{amsmath}
\usepackage{amssymb}
\usepackage{color}
\usepackage{graphicx}
\usepackage{comment}
\usepackage{multirow}
\usepackage[dvipsnames]{xcolor}

\usepackage[utf8]{inputenc}

\usepackage{caption}

\include{journals}

\graphicspath{{images/}}

\newcommand{\om}{$\Omega_\mathrm{M}$}

\def\gsim{\mathrel{\rlap{\lower 4pt \hbox{\hskip 1pt $\sim$}}\raise 1pt \hbox {$>$}}}
\def\lsim{\mathrel{\rlap{\lower 4pt \hbox{\hskip 1pt $\sim$}}\raise 1pt \hbox {$<$}}}

\begin{document}

\title{\boldmath The effect of inhomogeneities on dark energy constraints}

\author[a]{Suhail Dhawan,}
\affiliation[a]{Oskar Klein Centre, Department of Physics, Stockholm University, SE 106 91 Stockholm, Sweden}
\emailAdd{suhail.dhawan@fysik.su.se}

\author[a]{Ariel Goobar,}
\emailAdd{ariel@fysik.su.se}

\author[a]{Edvard M{\"o}rtsell}
\emailAdd{edvard@fysik.su.se}


\abstract{
Constraints on models of the late time acceleration of the universe assume the cosmological principle of homogeneity and isotropy on large scales. However, small scale inhomogeneities can alter observational and dynamical relations, affecting the inferred cosmological parameters. For precision constraints on the properties of dark energy, it is important to assess the potential systematic effects arising from these inhomogeneities. In this study, we use the Type Ia supernova  magnitude-redshift relation to constrain the inhomogeneities as described by the Dyer-Roeder distance relation and the effect they have on the dark energy equation of state ($w$), together with priors derived from the most recent results of the  measurements of the power spectrum of the Cosmic Microwave Background and Baryon Acoustic Oscillations. We find that the parameter describing the inhomogeneities ($\eta$) is weakly correlated with $w$. The best fit values $w = -0.933 \pm 0.065$ and $\eta = 0.61 \pm 0.37$  are consistent with homogeneity at $< 2 \sigma$ level. Assuming homogeneity ($\eta =1$), we find $w = -0.961 \pm 0.055$, indicating only a small change in $w$.  For a time-dependent dark energy equation of state, $w_0 = -0.951 \pm 0.112$ and $w_a = 0.059 \pm 0.418$, to be compared with $w_0 = -0.983 \pm 0.127$ and $w_a = 0.07 \pm 0.432$ in the homogeneous case, which is also a very small change. We do not obtain constraints on the fraction of dark matter in compact objects, $f_p$, at the 95$\%$ C.L. with conservative corrections to the distance formalism. Future supernova surveys will improve the constraints on $\eta$, and hence, $f_p$, by a factor of $\sim$ 10.  
}

\maketitle
\section{Introduction}
The discovery of the accelerated expansion of the universe  \cite{1998AJ....116.1009R,1999ApJ...517..565P} indicates the existence of a cosmological fluid with negative pressure, dark energy. However, there are still significant problems in the standard model describing the current epoch of accelerated expansion \cite[see][for a review]{2000astro.ph..5265W}. Current observations of Type Ia supernovae \cite[SNe~Ia;][]{2014A&A...568A..22B} the cosmic microwave background \cite[CMB;][]{2016A&A...594A..14P} and baryon acoustic oscillation \cite[BAO;]{2016arXiv160703155A} have constrained the dark energy equation of state (EoS; the ratio of the pressure to density $w = P/\rho \sim -1$  with $\lesssim$ 5$\%$ precision). Several models for cosmic acceleration  e.g. scalar fields \cite{1988ApJ...325L..17P,1988NuPhB.302..668W} or a modification of general relativity could be a viable explanation of the observations \cite[see][for a model comparison of the different scenarios]{2017JCAP...07..040D}.

The combination of SNe~Ia luminosity distances with BAO angular scale and CMB geometric distance priors provides the most stringent constraints on the properties of dark energy \cite{2016arXiv160703155A,2014A&A...568A..22B,2015PhRvD..92l3516A,2016A&A...594A..13P}.
 Apart from being crucial probes of the nature of dark energy, SNe~Ia are also important tracers of structure in the universe. They have been used to independently measure $\sigma_8$, the rms linear fluctuation in the matter distribution \cite{2014MNRAS.443L...6C}, with future surveys expected to significantly improve the precision on growth of perturbations \cite{2006PhRvD..74f3515D,2015MNRAS.449.2845A,2017MNRAS.465.2862S}.
Several  explanations of  the late time acceleration \cite[see][for a review]{2006IJMPD..15.1753C,2013CQGra..30u4003T,2016ARNPS..66...95J}, assume the cosmological principle, i.e. homogeneity and isotropy on large ($\sim$ 100 Mpc) scales, however, as the universe is known to be inhomogeneous on smaller scales, the matter distribution in the line of sight to SNe~Ia will effect the distance-redshift relation. An approximation was proposed to account for light traveling through emptier rather than denser regions of the universe in \cite{{1964SvA.....8...13Z}}, \cite{1969ApJ...155...89K} proposed a correction to the luminosity-redshift relation in homogeneous models and \cite{1973ApJ...180L..31D} derived a distance-redshift relation, known as the Dyer-Roeder (DR) relation. In this {\em ansatz}, the total matter density, $\rho_m$ contributes to the expansion of the universe, whereas a fraction of the density, $\eta \cdot \rho_m$,  contributes to the focusing of the light in the line of sight to the source. The contribution of shear from inhomogeneities in the magnification is assumed to be negligible. 
This approximate distance-redshift relation allows us to measure the inhomogeneity of the universe through , 1-$\eta$. The detection of gravitational waves (GW) from binary black hole mergers \cite{2016PhRvL.116f1102A,2016PhRvL.116x1103A,2017PhRvL.118v1101A} has renewed interest in compact objects comprising a fraction of the dark matter (DM) content of the universe \cite[see][for a review]{2016PhRvD..94h3504C} and hence, observational limits from the magnitude-redshift relation of SNe~Ia would be useful in testing this hypothesis. In the past, the study of the lensing distribution of SNe~Ia has been proposed as a viable way to shed new light into this issue \cite{2001ApJ...559...53M}, and the recent discovery of the  strongly lensed SN~Ia iPTF16geu \cite{2017Sci...356..291G} further emphasizes the need to explore this possibility.

In this paper, we analyse the effect of inhomogeneities on the inferred properties of dark energy (density, EoS and its time dependence). We use  constraints on  inhomogeneities to evaluate the fraction of dark matter (DM) in compact objects from current and future SN~Ia. 


In section~\ref{sec:data} we introduce the methodology and the data used in this work. In section~\ref{sec:par_est} we present the parameter constraints for the different cosmological model scenarios in our study which are discussed in Section~\ref{sec:disc}

\section{Methodology and data}
\label{sec:data}

\subsection{Type Ia supernovae}
For our analysis we use the most recent SN~Ia magnitude-redshift relation from the JLA compilation \citep{2014A&A...564A.125B}. 

Theoretically, the distance modulus predicted by the homogeneous and isotropic, flat Friedman-Robertson-Walker (FRW) universe is given by

\begin{equation}
\mu(z; \theta) = 5\, \mathrm{log_{10}} \left( \frac{d_L}{10\, \mathrm{Mpc}} \right) + 25
\label{eq:mu_sne}
\end{equation}
where $z$ is the redshift, $\theta$ are the cosmological parameters and $d_L$ is given by 

\begin{equation}
D_L = \frac{c(1+z)}{H_0 \sqrt{|\Omega_\mathrm{K}|}}\, \mathrm {sinn}\, \left( \sqrt{|\Omega_\mathrm{K}|} \int^{z}_{0} \frac{dz^{'}}{E(z^{'})} \right)
\label{eq:lum_dist}
\end{equation}
where $\Omega_K$ is the dimensionless curvature density and $\mathrm{sinn} (x) = \{ \mathrm{sin}(x), x, \mathrm{sinh}(x) \}$ for close, flat and open universes, respectively. $E(z) = H(z)/H_0$ is the normalised Hubble parameter.

\begin{equation}
E(z)^2   = \Omega_\mathrm{M} (1+z)^3 + \Omega_{\mathrm{DE}}(z) + \Omega_\mathrm{K}(1+z)^2, 
\label{eq:norm_hubbleparameter}
\end{equation}
where 

\begin{equation}
\mathrm{\Omega_{\mathrm{DE}}(z)} = \Omega_{\mathrm{DE}}\, \mathrm{exp} \left[ 3 \int_0^z \frac{1+w(x)}{1 + x} dx \right],
\label{eq:ode_z}
\end{equation}
$w(z)$ is the dark energy EoS. 

Observationally, the distance modulus is calculated from the SN~Ia peak apparent magnitude ($m_B$), light curve width ($x_1$) and colour ($c$)

\begin{equation}
\mu_{obs} = m_B - (M_B - \alpha x_1 + \beta c),
\label{eq:obs_distmod}
\end{equation}
where $M_B$ is the absolute magnitude of the SN~Ia. Following \cite{2014A&A...568A..22B}, we apply a step correction ($\Delta_\mathrm{M}$) for the host galaxy stellar mass. We note that $\alpha$, $\beta$, $M_B$ and $\Delta_\mathrm{M}$ are nuisance parameters in the fit for the cosmology. 

The $\chi^2$ is given by 
\begin{equation}
\chi_{\mathrm{SN}}^2 = \Delta^T C_{\mathrm{SN}}^{-1} \Delta, 
\end{equation}
where $\Delta = \mu - \mu_{obs}$ and $C$ is the complete covariance matrix described in \cite{2014A&A...568A..22B}.

\subsection{Cosmic Microwave Background}
Precise constraints on the expansion history can be derived by combining the SN~Ia magnitude-redshift relation with complementary cosmological probes \cite{2006astro.ph..9591A}. 
For the geometric constraints from the CMB we use the compressed likelihood from the \emph{Planck} satellite,  marginalised over the lensing amplitude ($A_L$) \cite[see][for details]{2016A&A...594A..14P}. The CMB shift, $R$, position of the first acoustic peak in the power spectrum, $l_A$ and the baryon density at present day, $\Omega_b h^2$ comprise the data vector. The expression for the CMB shift and the position of the first acoustic peak are given by

\begin{equation}
R = \sqrt{\Omega_\mathrm{M} H_0^2} d_A(z_*)/c,
\label{eq:cmb_shift}
\end{equation}
and
\begin{equation}
l_A = \pi \frac{d_A(z_*)}{r_s(z_*)}
\label{eq:cmb_la}
\end{equation}
where $r_s(z)$ is the sound horizon at redshift, $z$, given by 
\begin{equation}
r_s(z) = \frac{c}{\sqrt{3}} \int^{\frac{1}{1+z}}_0 \frac{da}{a^2 H(a) \sqrt(1 + a\frac{3\Omega_\mathrm{b}}{4\Omega_\gamma})}, 
\label{eq:sound_horizon}
\end{equation}
where $\Omega_\gamma = 2.469\cdot 10^{-5} h^{-2}$ and $h= H_0/100$ \cite[see][for more details]{2009ApJS..180..330K}.

The value for ($R$,$l_A$, $\Omega_b h^2$) = (1.7382, 301.63, 0.02262) with errors (0.0088,0.15,0.00029) and covariance is
\begin{equation}
 D_\text{CMB} = \left(
\begin{array}{ccc}
1.0 &  0.64 &  -0.75 \\
0.64 & 1.0 & -0.55 \\
 -0.75 & -0.55 & 1.0 \\
\end{array}
\right)\,,
\label{eq:cmb_covar_pl15}
\end{equation}
such that the elements of the covariance matrix $C_{ij} = \sigma_i \sigma_j D_{ij}$

\begin{figure}
\includegraphics[width=.52\textwidth]{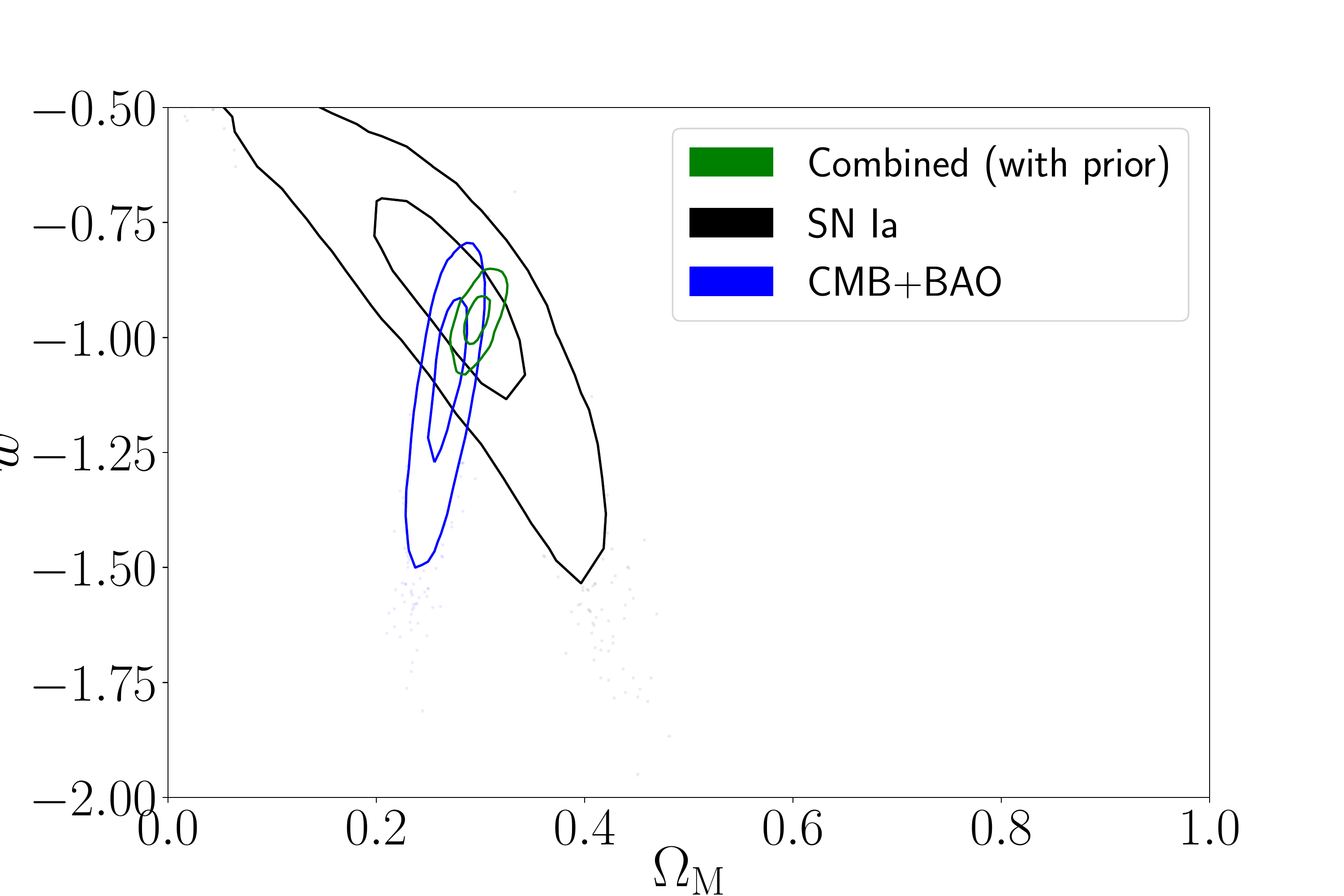}
\includegraphics[width=.52\textwidth]{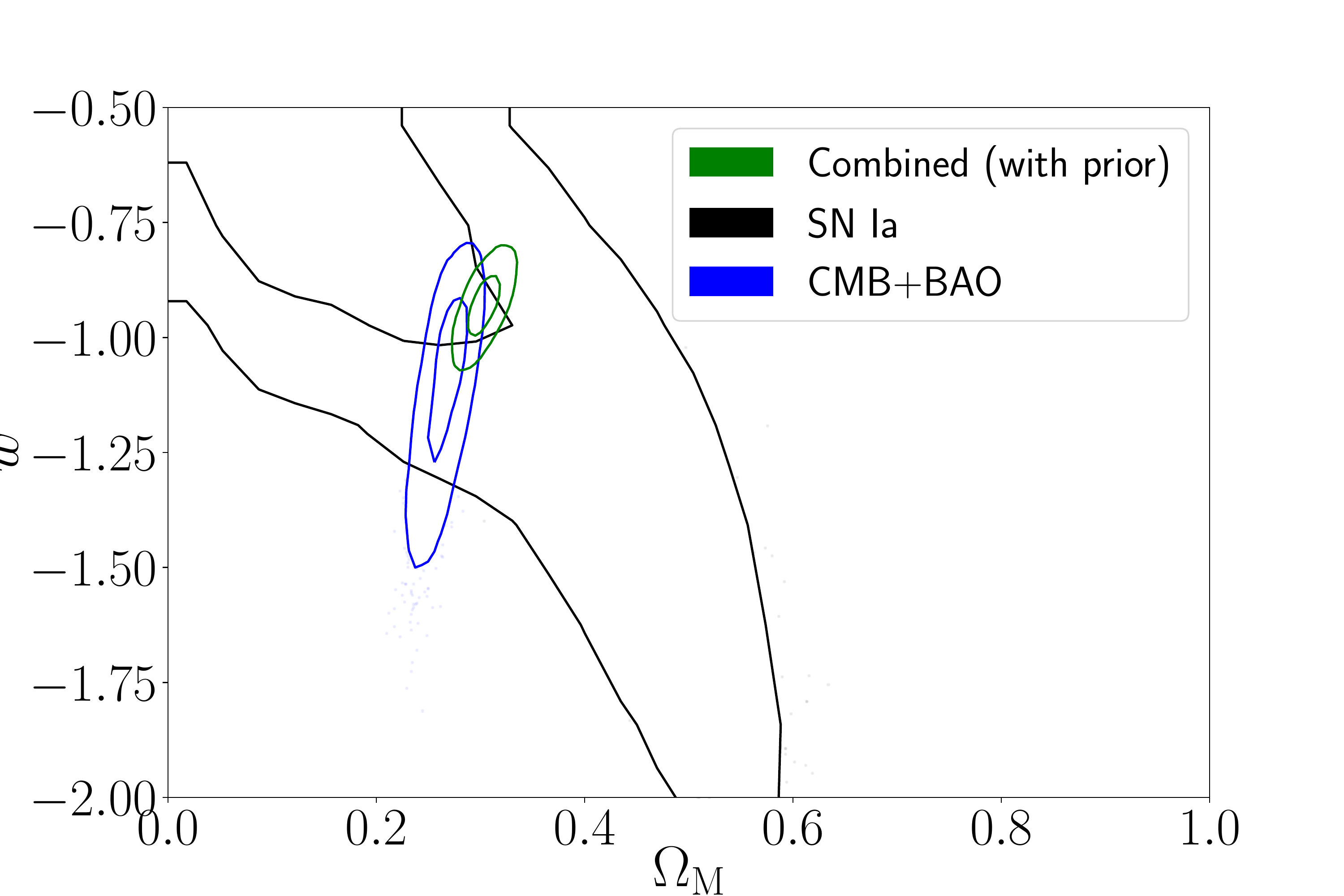}
\caption{Joint constraints on the present day matter density $\Omega_\mathrm{M}$ and the equation of state of dark energy, $w$, under the assumption of homogeneity (left) and with $\eta$ as a free parameter (right). Although the constraints from the SN~Ia are degraded, combining them with the CMB/BAO prior yields very similar results in both cases. The contours are at 1 and 2 $\sigma$ level. The plot was made using the \texttt{python} package \texttt{corner} \cite{corner}.}
\label{fig:w_omegam_eta}
\end{figure}

\subsection{Baryon Acoustic Oscillation}
\label{ssec:bao}

The detection of the characteristic scale of the BAO in the correlation function of different matter distribution tracers provides a powerful standard ruler to probe the angular-diameter-distance versus redshift relation. BAO analyses usually perform a spherical average  constraining a combination of the angular scale and redshift separation

\begin{equation}
d_z = \frac{r_s(z_{\mathrm{drag}})}{D_V(z)}, 
\label{eq:dzbao}
\end{equation}
with 
\begin{equation}
D_V(z) = \left[(1+z)^2 D_A(z)^2 \frac{cz}{H(z)} \right] ^{1/3}, 
\label{eq:dvz}
\end{equation}
where $D_A$ is the angular diameter distance. 
$r_s(z_{\mathrm{drag}})$, is the sound horizon at the drag redshift given by Equation~\ref{eq:sound_horizon}.

For our analyses, we follow the method of \cite{2014A&A...568A..22B} and use three measurements  at  $z_{\mathrm{eff}}$ = 0.106, 0.35 and 0.57 from \cite{2011MNRAS.416.3017B,2012MNRAS.427.2132P,2012MNRAS.427.3435A} respectively.
We consider a BAO prior of the form 
\begin{equation}
\chi^2_{\mathrm{BAO}} = (d_z - d_z^{\mathrm{BAO}})^T C_{\mathrm{BAO}}^{-1} (d_z - d_z^{\mathrm{BAO}}), 
\label{eq:chi_bao}
\end{equation}
with $d_z^{\mathrm{BAO}}$ = [0.336, 0.1126, 0.07315] and $C_{\mathrm{BAO}}^{-1}$ = diag(4444., 215156., 721487.). 
        
We note that the WiggleZ team also presents three distance measurements  \cite[see][]{2014MNRAS.441.3524K,2011MNRAS.416.3017B}. However, since the WiggleZ volume partially overlaps with that of the BOSS CMASS sample, and the correlations have not been quantified, we do not include the WiggleZ results in this study.

\section{Parameter Estimation}
\label{sec:par_est}
In this section, we quantify the parameter constraints on the dark energy EoS when using the DR distance redshift relation. We also compute the Hubble residuals for the SNe~Ia and compare the skewness in each redshift bin to simulations with varying fractions of dark matter in compact objects. 

\subsection{Effect on dark energy constraints}
\label{ssec:DE}
The general differential equation for the distance between two light rays of the boundary of a small light cone propagating far away from all clumps of matter in an inhomogenous universe is developed in \cite{1964SvA.....8...13Z, 1965SvA.....8..854D,1966SvA.....9..671D,1966RSPSA.294..195B,1967ApJ...150..737G,1992grle.book.....S}. This relation for angular diameter distance as a function of the inhomogeneities given in \cite{1973ApJ...180L..31D} is

\begin{equation}
Q D^{\prime\prime} + (\frac{2Q}{1+z} + \frac{Q^{\prime}}{2}) D^{\prime} + \frac{3}{2} \eta\, \Omega_\mathrm{M}\, (1+z)\, D = 0 ,
\label{eq:dr_dist}
\end{equation}
where $\eta$ is the fraction of the matter density in opaque clumps and $Q(z) = E(z)^2$, hence, $Q(z)$ depends on the cosmological model. In previous studies 
\cite[e.g.][]{2013PhRvD..87f3527B,2015MNRAS.451.2097H}, the model has been fixed to the standard $\Lambda$CDM cosmology, assuming that the accelerated expansion is caused by a cosmological constant. In some cases, even the background cosmological parameters, e.g. $\Omega_\mathrm{M}$, $\Omega_\Lambda$ have been fixed to the values from concordance cosmology \cite{2015MNRAS.451.2097H}. 

\begin{figure}
\centering
\includegraphics[width=.8\textwidth, trim = 0 0 0 30]{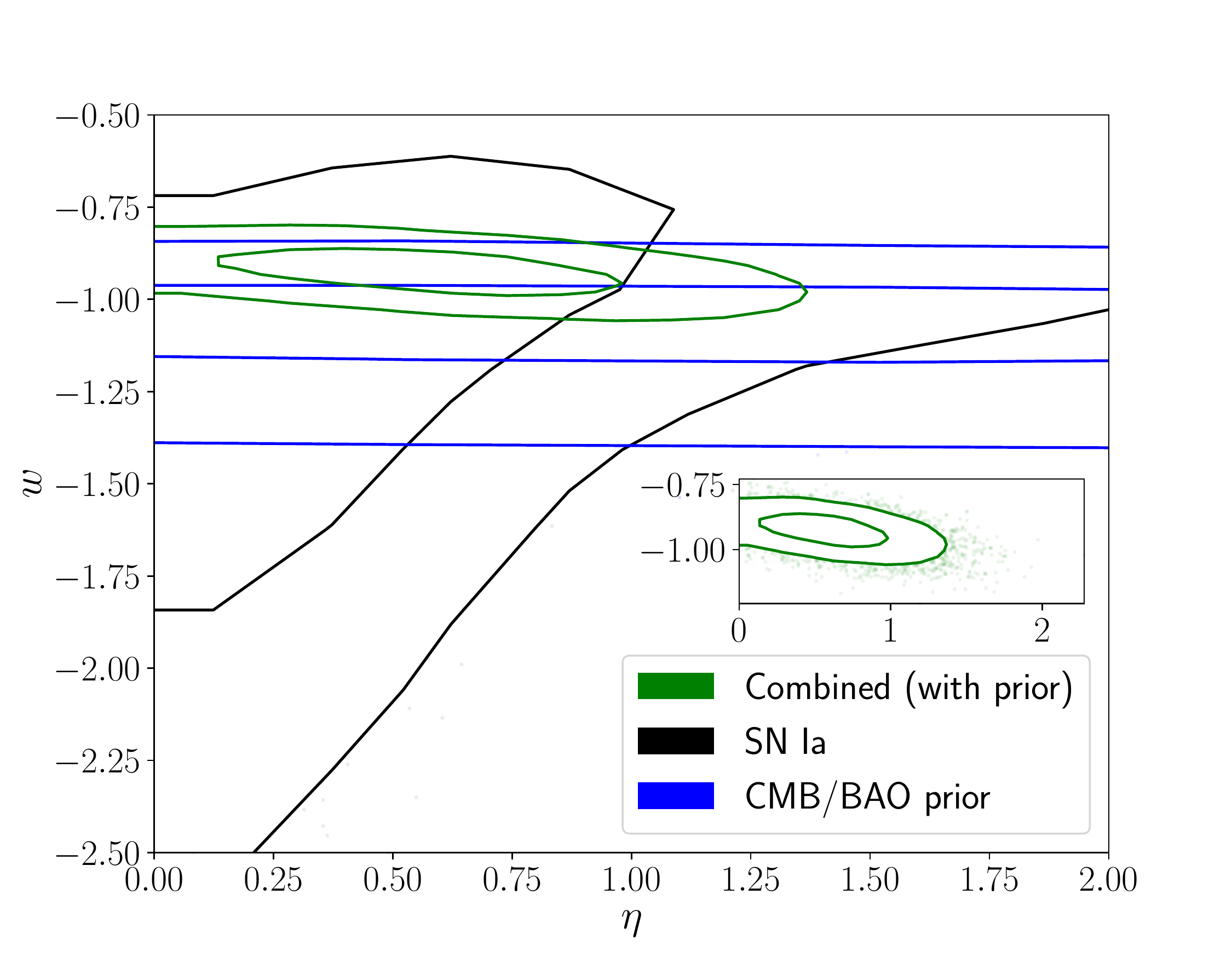}
\caption{Joint constraints on $w$ and $\eta$ from SNe~Ia (black) with a prior on $w$CDM cosmology from the CMB and BAO data (blue). The combined constraints are plotted in green in the inset. The contours are at the 1 and 2 $\sigma$ level.  }
\label{fig:dr_above1}
\end{figure}

Here, we evaluate the constraints on dark energy EoS along with the parameter describing the small scale inhomogeneity of the universe. We use the  expression for $Q(z)$ for general dark energy models 
\begin{equation}
Q(z) = \Omega_\mathrm{M} (1+z)^3 + \Omega_\mathrm{K} (1+z)^2 + \Omega_{\mathrm{DE}}\,(z, w)
\label{eq:qz}
\end{equation}
with $w$, the EoS of dark energy being a free parameter. We test two simple cases, a constant $w$ and time-varying parametrisation $w(a) = w_0 + w_a (1-a)$ \cite{2001IJMPD..10..213C,2003PhRvL..90i1301L}, however, this can be extended to any expression for $w(a)$. We assume flatness, i.e. $\Omega_\mathrm{M} + \Omega_{\mathrm{DE}} = 1$.

We fit the SN~Ia Hubble diagram with the luminosity distance which is derived from the angular diameter distance using the distance duality relation \cite{1933PMag...15..761E,2004PhRvD..69j1305B,2011JCAP...05..023N,2016arXiv161109426H}. 

\begin{equation}
D_L = D_A (1+z)^2.
\label{eq:ddr}
\end{equation}
The lower limit on the prior for $\eta$ is set by $\eta > 0$. If there are no small scale inhomogeneities, then $\eta = 1$. Finding $\eta > 1$ is possible if the average density in the beam is larger than the global density (such cases could be result of a selection effect). 

\begin{table*}
\caption{Priors on the free parameters for the DR angular diameter distance. The four parameters listed here, namely, $\alpha$, $\beta$, $\Delta_M$, $M_B$ are nuisance parameters that need to fit to the supernova data. We use a flat prior on $H_0$.}
\begin{center}
\begin{minipage}{70mm}
\begin{tabular}{|l|c|}
\hline\hline
Parameter & Prior \\
\hline
\om & U[0, 1] \\
$\eta$\footnote{In specific cases the prior  on $\eta$ is U[0, 1]} & U[0, 5.]\\
$w$ & U[-2, 2] \\
$H_0$ & U[50, 100] \\
&\\

$\alpha$ & U[0, 1] \\
$\beta$ & U[0, 4] \\
$\Delta_M$ & U[0, 0.2] \\
$M_B$ & U[-35, 15] \\
\hline
\end{tabular}
\end{minipage}
\end{center}
\label{tab:prior}
\end{table*}

The inhomogeneities investigated in this work are on scales much smaller than probed by the CMB and BAO observations \cite{1976ApJ...208L...1W,2012MNRAS.426.1121C}, hence, these data are fit with the assumption of homogeneity (i.e. $\eta = 1$). 
DM in compact objects affects the distances but not the expansion history \cite[however, see backreaction models, e.g.][]{2017arXiv170906022M}.  Therefore, the only probe sensitive to $\eta$ is SN~Ia. The masses considered here given by  a comparison of Einstein radii and size of SNe \cite{2017Sci...356..291G}, corresponding to $\gtrsim$ 10$^{-2}$ $M_{\odot}$ \cite{1999A&A...351L..10S}.

We fit Equation~\ref{eq:dr_dist} to the SN~Ia Hubble diagram. The complementary probes help to constrain the background cosmological parameters. We impose a conservative, uniform prior of $\eta$ (see Table~\ref{tab:prior}). $\eta > 1$ corresponds to the average density in the beam being larger than the global density, but we do not strictly impose $\eta < 1$ at this stage of the inference. We constrain $\eta$ along with the parameters describing the dark energy density, equation of state, and its time-dependence. 

\subsubsection{Constant $w$}
We infer $w = -0.933 \pm 0.065$ which is consistent with the cosmological constant and $\eta = 0.61 \pm 0.37$ consistent with homogeneity within the statistical accuracy of this study.

To investigate the effect of  inhomogeneities on the inferred $w$, we fit all the datasets with $\eta =1$ i.e. no inhomogeneities and find $w = -0.961 \pm 0.055$ slightly shifted from the case where $\eta$ is a free parameter. We also fit for $\eta$ fixing $w= -1$ i.e. concordance cosmology (red histogram, Figure~\ref{fig:dr_above1}) and infer $\eta = 0.81 \pm 0.33$. Using the relation between $\eta$ and the fraction of DM in compact objects, $f_p$, from \cite{2002A&A...382..787M} we get  $f_p < 0.73$ at 68$\%$ C.L., however, we cannot constrain $f_p$ at the 95$\%$ C.L.  We note that without using the relation from the \texttt{SNOC} simulations, that conservatively corrects for Weyl focussing, we get a limit of $f_p < 0.81$ at the 95$\%$ C.L.

We note that the shift in $w$ in the case with inhomogeneities allowed is very small (0.03) and that the increase in the uncertainty is small. In both the $\Lambda$CDM and $w$CDM background cases, the value of $\eta$ is consistent with homogeneity.

\begin{figure}
\centering
\includegraphics[width=.8\textwidth, trim = 0 0 30 0]{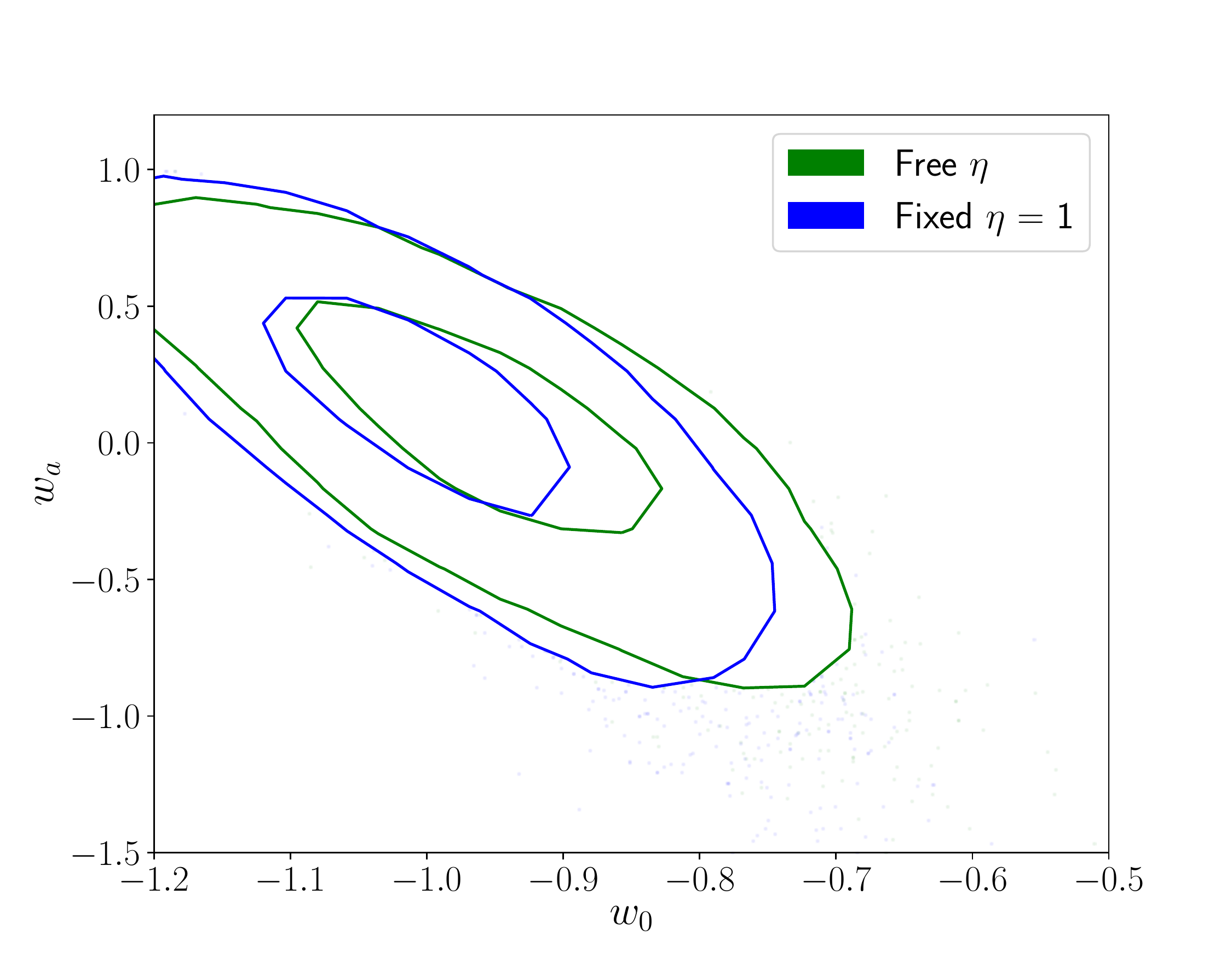}
\caption{Constraints on $w_0$ and $w_a$ (see Section~\ref{sssec:w0wa}). The blue contours are the constraints with $\eta = 1$. The contours are at the 1 and 2 $\sigma$ level}
\label{fig:w0_wa}
\end{figure}

\subsubsection{Time-Dependent Equation of state}
\label{sssec:w0wa}
We note that in the constant $w$ model, there is only a small effect of $\eta$ being a free parameter. We test a general time-dependent equation of state parametrisation of $w(a) = w_0 + w_a (1 - a)$ and present constraints in Figure~\ref{fig:w0_wa}.

We find  $w_0 = -0.954 \pm 0.112$ and $w_a = 0.059 \pm 0.418$  consistent with a cosmological constant,  $\eta = 0.67 \pm 0.39$ consistent with homogeneity. Fixing $\eta = 1$ $w_0 = -0.982 \pm 0.127$ and $w_a = 0.070 \pm 0.432$. The change in the $w_0$ and $w_a$ is very small with a slight increase in the uncertainties.

\subsection{Hubble residuals}
\label{ssec:hubble_res}
A complementary method we employ to test for the presence of inhomogeneities involves a direct comparison of the observed SN~Ia Hubble residuals and expected perturbations for a given fraction of DM in compact objects, $f_p$ \cite{2007PhRvL..98g1302M}. For future surveys, forecasts for $f_p$ were evaluated in \cite{2001ApJ...559...53M}. 
We simulate the  perturbations using  the ray-tracing software \texttt{SNOC} \cite{2002A&A...392..757G} for 8 different values of $f_p$, namely 0, 0.05, 0.10, 0.15, 0.20, 0.25, 0.50, 0.80, 1. 

We calculate the Hubble residuals using an input $\Lambda$CDM universe with $\Omega_M = 0.3$ and take the nuisance parameters from \cite{2014A&A...564A.125B}, i.e. $\alpha = 0.141$, $\beta = 3.102$. We use only the Supernova Legacy Survey (SNLS) subsample \cite{2011ApJS..192....1C,2011ApJ...737..102S} of the JLA compilation since it covers a wide range of redshifts and is a uniformly observed set of SNe providing a ``clean" dataset to constrain the value of $f_p$. For this subsample, we use an intrinsic scatter of 0.08 mag, as reported in \cite{2014A&A...564A.125B}.

We compute 10 000 Monte Carlo realisations from a distribution $\mathcal{N} (0, \sigma_i)$ where $\sigma_i$ is the measurement error for the $i^{th}$ SN which includes  the error in $m_B$, $x_1$ and $c$ as well as the intrinsic scatter for the SNLS subsample. Each iteration of the realisations is added to the perturbations calculated from \texttt{SNOC} and a Kolmogorov-Smirnov (KS) test is used to determine whether both the  realisations and the Hubble residuals are drawn from the same parent population, usingjca the criterion that $p < 0.05$ suggests that they come from different parent populations. A value of $f_p$ can be excluded if more than 90$\%$ of the realisations are seen to be drawn from a different parent population from the Hubble residuals.

Applying the above test to the $f_p$ values from 0 to 1, we find that none of the simulations have a significant fraction of the realisations that appear to be from a different parent population than the residuals (for the most extreme case of $f_p = 1$ only $\sim$ 60$\%$ of the realisations have a $p$-value $< 0.05$). Hence, all of the $f_p$ values are consistent with the data and we cannot draw any conclusions on the fraction of DM in compact objects by just analysing the Hubble residuals of the SNe~Ia. 

We note that we do not obtain stringent constraints on $f_p$ compared to recent studies, \cite[e.g.][]{2017arXiv171202240Z} since the authors use the complete JLA catalog with the highest-$z$ supernovae while we restrict our analyses to the largest uniform measured subsample since we conservatively don't want to be affected by selection effects and photometric systematics.	 We note that \cite{2018PDU....20...95G} find  fraction being $< 1.09$ at 95$\%$ C.L.,  which is similar to the constraints presented here. This is because their assumptions on the compact object mass distribution, SN physical size and background cosmology differ from the ones in  \cite{2017arXiv171202240Z}.
\begin{figure}
\centering
\includegraphics[width=.9\textwidth]{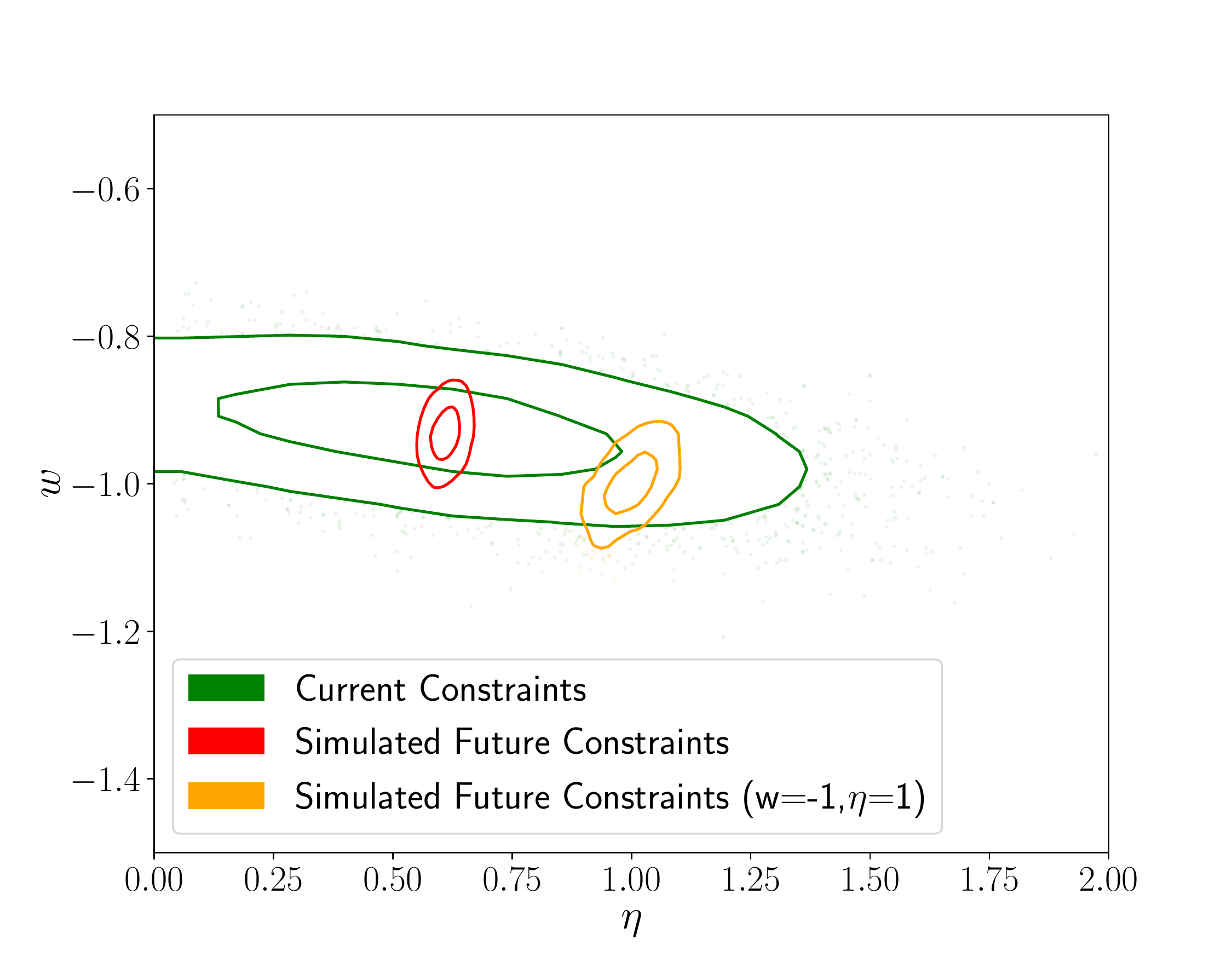}
\caption{A comparison of the current constraints on $w$-$\eta$ (green) with forecasts (red) from a future SN~Ia experiment \cite{2014A&A...572A..80A}. Using the best fit parameters from current fits we find that we can exclude $\eta$ =1 significantly ($\sigma(\eta)$ = 0.03). We also show the constraints for an underlying cosmology with $w =-1$ and $\eta$=1, which will constrain the fraction of DM in compact objects, $f_p < 12 \%$ at 95$\%$ C.L.  The contours are at the 1 and 2 $\sigma$ level.}
\label{fig:forecast}
\end{figure}

\section{Forecast for future SN survey}
Although current constraints are consistent with homogeneity, the limits on the fraction of DM in compact objects are weak. Here, we investigate  improvements in parameter estimates on $\eta$ and $w$ with future SN~Ia surveys. 

For our analyses we use the forecasts from the DESIRE survey \cite{2014A&A...572A..80A} which combines a ground-based low-$z$ and intermediate-$z$ with LSST with a space-based high-$z$ lever extending to $\sim$ 1.5. We combine the future SN~Ia distances with current constraints from the CMB, BAO and $H(z)$ data (see Section ~\ref{sec:data}). We test cosmological with input parameters ($\eta$, $w$) = (0.61,-0.933), best fit values from current data and $( 1, -1)$, a homogenous universe with a cosmological constant. The resulting contours are plotted in Figure~\ref{fig:forecast}.

We obtain $\eta = 0.61 \pm 0.030$ and $\eta = 1.00 \pm 0.036$ (for the two input cases of $\eta=0.61$ and $\eta = 1$) which is a significant improvement on the current constraints (Figure~\ref{fig:forecast} has a comparison with current $w$CDM constraints). This would translate to a limit of $< 12 \%$ of DM in compact objects. We also note that constraints on $w$ improve to 4$\%$ compared to $\sim$ 7 $\%$ with current data (note that the CMB and BAO  data are the same as in the current analyses, hence, the improvement in $w$ is only from the SNe and not the complementary probes). The forecast illustrates the power of an improved SN~Ia Hubble diagram, beyond constraining dark energy models \cite[e.g.][]{2013arXiv1305.5422S,2014A&A...572A..80A}, to test homogeneity and precisely estimate the fraction of DM in compact objects. We obtain similar limits on $\eta$ ($\sim$ 5$\%$) from the expected Hubble diagram of SN~Ia observed with the WFIRST satellite \cite{2013arXiv1305.5422S}. 

We also constrain time-dependent dark energy EoS in the presence of inhomogeneities. We find that there is very little effect of $\eta$ as a free parameter on the inferred values of $w_0$ and $w_a$ which are $-0.954 \pm 0.113$ and $0.05 \pm 0.418$ which is consistent with previous analyses \cite[e.g.][]{2016arXiv160703155A}.  
Hence, for various dark energy properties, namely, the density, EoS and time dependence, the presence of possible inhomogeneities does not significantly change the parameter constraints.

\section{Discussion and Conclusion}
\label{sec:disc}

We studied the effect of  non-uniformities on inferred properties of dark energy from the SNe~Ia Hubble diagram. The impact of inhomogeneities on dark energy parameters was theoretically studied in \cite{2012JCAP...11..045B,2013PhRvL.110b1301B,2017JCAP...03..062F} and of ``Swiss-Cheese" cosmological models in \cite{2013PhRvD..87l3526F,2014PhRvD..90l3536P,2014JCAP...06..054F}.  The cumulative effect of the small scale inhomogeneities could (completely, or, in part) mimick a cosmological constant by altering the observational and dynamical relation on large scales \cite{2009JPhCS.189a2011E}. Hence, it is critical to evaluate the effect of the clumpiness parameter, $\eta$, on the inference of $w$. We find that $\eta$ is degenerate with $w$, such that a higher value of $\eta$ implies a more negative $w$. The combination of SNe~Ia, BAO, CMB constrains $w = -0.933 \pm 0.065$. Both for a fixed $\eta = 1$, and with $\eta$ as a free parameter  $w$ is consistent at  $< 2 \sigma$ of the cosmological constant. 

For a background concordance cosmology (i.e. $w = -1$), we confirm previous analyses \cite{2013JCAP...06..007Y,2015MNRAS.451.2097H} that find $\eta = 0.81 \pm 0.33$ at the 68$\%$ C.L.  However, we stress that we do not a priori fix the cosmological parameter values as in \cite{2015MNRAS.451.2097H}, but rather we constrain it with data complementary to SN~Ia distances. We also obtain more stringent constraints than \cite{2015PhRvD..91h3010L} since we use complementary datasets that constrain $\Omega_M$ very well. Using the relation between $\eta$ and the fraction of DM in compact objects ($f_p$) \cite{2002A&A...382..787M} we find $f_p < 0.73$ at 68$\%$ C.L., but the data are not constraining at the 95$\%$ C.L.   However, we note the relation between $\eta$ and $f_p$ accounts for the Weyl focussing from matter outside the beam. If we do not make this conservative correction we can constrain $f_p < 0.81$ at the 95$\%$ C.L. Some recent studies using less conservative constraints \cite[e.g.][]{2017arXiv171202240Z} find stronger constraints on $f_p$, however \cite{2018PDU....20...95G} , under another very different set of assumption, find the data to be consistent with all DM in compact objects.

Future SNe~Ia, {that aim to significantly increase the number of SNe at $z > 1$}, would be crucial to precisely estimate the fraction of DM in compact objects and therefore, evaluate departures from homogeneity.

{\it Acknowledgements:} 
This work was funded through grants from the Knut and Alice Wallenberg foundation, the Swedish Research Council (VR) and the Swedish National Space Board. We would like to thank Rahul Biswas and Aoife Boyle for useful discussions.

\bibliographystyle{unsrt}
\bibliography{biblio,SNIa}

\begin{thebibliography}{10}

\bibitem{1998AJ....116.1009R}
A.~G. {Riess}, A.~V. {Filippenko}, P.~{Challis}, A.~{Clocchiatti},
  A.~{Diercks}, P.~M. {Garnavich}, R.~L. {Gilliland}, C.~J. {Hogan}, S.~{Jha},
  R.~P. {Kirshner}, B.~{Leibundgut}, M.~M. {Phillips}, D.~{Reiss}, B.~P.
  {Schmidt}, R.~A. {Schommer}, R.~C. {Smith}, J.~{Spyromilio}, C.~{Stubbs},
  N.~B. {Suntzeff}, and J.~{Tonry}.
\newblock {Observational Evidence from Supernovae for an Accelerating Universe
  and a Cosmological Constant}.
\newblock {\em \aj}, 116:1009--1038, September 1998.

\bibitem{1999ApJ...517..565P}
S.~{Perlmutter}, G.~{Aldering}, G.~{Goldhaber}, R.~A. {Knop}, P.~{Nugent},
  P.~G. {Castro}, S.~{Deustua}, S.~{Fabbro}, A.~{Goobar}, D.~E. {Groom}, I.~M.
  {Hook}, A.~G. {Kim}, M.~Y. {Kim}, J.~C. {Lee}, N.~J. {Nunes}, R.~{Pain},
  C.~R. {Pennypacker}, R.~{Quimby}, C.~{Lidman}, R.~S. {Ellis}, M.~{Irwin},
  R.~G. {McMahon}, P.~{Ruiz-Lapuente}, N.~{Walton}, B.~{Schaefer}, B.~J.
  {Boyle}, A.~V. {Filippenko}, T.~{Matheson}, A.~S. {Fruchter}, N.~{Panagia},
  H.~J.~M. {Newberg}, W.~J. {Couch}, and T.~S.~C. {Project}.
\newblock {Measurements of {$\Omega$} and {$\Lambda$} from 42 High-Redshift
  Supernovae}.
\newblock {\em \apj}, 517:565--586, June 1999.

\bibitem{2000astro.ph..5265W}
S.~{Weinberg}.
\newblock {The Cosmological Constant Problems (Talk given at Dark Matter 2000,
  February, 2000)}.
\newblock {\em ArXiv Astrophysics e-prints}, May 2000.

\bibitem{2014A&A...568A..22B}
M.~{Betoule}, R.~{Kessler}, J.~{Guy}, J.~{Mosher}, D.~{Hardin}, R.~{Biswas},
  P.~{Astier}, P.~{El-Hage}, M.~{Konig}, S.~{Kuhlmann}, J.~{Marriner},
  R.~{Pain}, N.~{Regnault}, C.~{Balland}, B.~A. {Bassett}, P.~J. {Brown},
  H.~{Campbell}, R.~G. {Carlberg}, F.~{Cellier-Holzem}, D.~{Cinabro},
  A.~{Conley}, C.~B. {D'Andrea}, D.~L. {DePoy}, M.~{Doi}, R.~S. {Ellis},
  S.~{Fabbro}, A.~V. {Filippenko}, R.~J. {Foley}, J.~A. {Frieman},
  D.~{Fouchez}, L.~{Galbany}, A.~{Goobar}, R.~R. {Gupta}, G.~J. {Hill},
  R.~{Hlozek}, C.~J. {Hogan}, I.~M. {Hook}, D.~A. {Howell}, S.~W. {Jha}, L.~{Le
  Guillou}, G.~{Leloudas}, C.~{Lidman}, J.~L. {Marshall}, A.~{M{\"o}ller},
  A.~M. {Mour{\~a}o}, J.~{Neveu}, R.~{Nichol}, M.~D. {Olmstead},
  N.~{Palanque-Delabrouille}, S.~{Perlmutter}, J.~L. {Prieto}, C.~J.
  {Pritchet}, M.~{Richmond}, A.~G. {Riess}, V.~{Ruhlmann-Kleider}, M.~{Sako},
  K.~{Schahmaneche}, D.~P. {Schneider}, M.~{Smith}, J.~{Sollerman},
  M.~{Sullivan}, N.~A. {Walton}, and C.~J. {Wheeler}.
\newblock {Improved cosmological constraints from a joint analysis of the
  SDSS-II and SNLS supernova samples}.
\newblock {\em \aap}, 568:A22, August 2014.

\bibitem{2016A&A...594A..14P}
{Planck Collaboration}, P.~A.~R. {Ade}, N.~{Aghanim}, M.~{Arnaud},
  M.~{Ashdown}, J.~{Aumont}, C.~{Baccigalupi}, A.~J. {Banday}, R.~B.
  {Barreiro}, N.~{Bartolo}, and et~al.
\newblock {Planck 2015 results. XIV. Dark energy and modified gravity}.
\newblock {\em \aap}, 594:A14, September 2016.

\bibitem{2016arXiv160703155A}
S.~{Alam}, M.~{Ata}, S.~{Bailey}, F.~{Beutler}, D.~{Bizyaev}, J.~A. {Blazek},
  A.~S. {Bolton}, J.~R. {Brownstein}, A.~{Burden}, C.-H. {Chuang},
  J.~{Comparat}, A.~J. {Cuesta}, K.~S. {Dawson}, D.~J. {Eisenstein},
  S.~{Escoffier}, H.~{Gil-Mar{\'{\i}}n}, J.~N. {Grieb}, N.~{Hand}, S.~{Ho},
  K.~{Kinemuchi}, D.~{Kirkby}, F.~{Kitaura}, E.~{Malanushenko},
  V.~{Malanushenko}, C.~{Maraston}, C.~K. {McBride}, R.~C. {Nichol}, M.~D.
  {Olmstead}, D.~{Oravetz}, N.~{Padmanabhan}, N.~{Palanque-Delabrouille},
  K.~{Pan}, M.~{Pellejero-Ibanez}, W.~J. {Percival}, P.~{Petitjean},
  F.~{Prada}, A.~M. {Price-Whelan}, B.~A. {Reid}, S.~A.
  {Rodr{\'{\i}}guez-Torres}, N.~A. {Roe}, A.~J. {Ross}, N.~P. {Ross},
  G.~{Rossi}, J.~A. {Rubi{\~n}o-Mart{\'{\i}}n}, A.~G. {S{\'a}nchez},
  S.~{Saito}, S.~{Salazar-Albornoz}, L.~{Samushia}, S.~{Satpathy}, C.~G.
  {Sc{\'o}ccola}, D.~J. {Schlegel}, D.~P. {Schneider}, H.-J. {Seo},
  A.~{Simmons}, A.~{Slosar}, M.~A. {Strauss}, M.~E.~C. {Swanson}, D.~{Thomas},
  J.~L. {Tinker}, R.~{Tojeiro}, M.~{Vargas Maga{\~n}a}, J.~A. {Vazquez},
  L.~{Verde}, D.~A. {Wake}, Y.~{Wang}, D.~H. {Weinberg}, M.~{White}, W.~M.
  {Wood-Vasey}, C.~{Y{\`e}che}, I.~{Zehavi}, Z.~{Zhai}, and G.-B. {Zhao}.
\newblock {The clustering of galaxies in the completed SDSS-III Baryon
  Oscillation Spectroscopic Survey: cosmological analysis of the DR12 galaxy
  sample}.
\newblock {\em ArXiv e-prints}, July 2016.

\bibitem{1988ApJ...325L..17P}
P.~J.~E. {Peebles} and B.~{Ratra}.
\newblock {Cosmology with a time-variable cosmological 'constant'}.
\newblock {\em \apjl}, 325:L17--L20, February 1988.

\bibitem{1988NuPhB.302..668W}
C.~{Wetterich}.
\newblock {Cosmology and the fate of dilatation symmetry}.
\newblock {\em Nuclear Physics B}, 302:668--696, June 1988.

\bibitem{2017JCAP...07..040D}
S.~{Dhawan}, A.~{Goobar}, E.~{M{\"o}rtsell}, R.~{Amanullah}, and U.~{Feindt}.
\newblock {Narrowing down the possible explanations of cosmic acceleration with
  geometric probes}.
\newblock {\em \jcap}, 7:040, July 2017.

\bibitem{2015PhRvD..92l3516A}
{\'E}.~{Aubourg}, S.~{Bailey}, J.~E. {Bautista}, F.~{Beutler}, V.~{Bhardwaj},
  D.~{Bizyaev}, M.~{Blanton}, M.~{Blomqvist}, A.~S. {Bolton}, J.~{Bovy},
  H.~{Brewington}, J.~{Brinkmann}, J.~R. {Brownstein}, A.~{Burden}, N.~G.
  {Busca}, W.~{Carithers}, C.-H. {Chuang}, J.~{Comparat}, R.~A.~C. {Croft},
  A.~J. {Cuesta}, K.~S. {Dawson}, T.~{Delubac}, D.~J. {Eisenstein},
  A.~{Font-Ribera}, J.~{Ge}, J.-M. {Le Goff}, S.~G.~A. {Gontcho}, J.~R. {Gott},
  J.~E. {Gunn}, H.~{Guo}, J.~{Guy}, J.-C. {Hamilton}, S.~{Ho}, K.~{Honscheid},
  C.~{Howlett}, D.~{Kirkby}, F.~S. {Kitaura}, J.-P. {Kneib}, K.-G. {Lee},
  D.~{Long}, R.~H. {Lupton}, M.~V. {Maga{\~n}a}, V.~{Malanushenko},
  E.~{Malanushenko}, M.~{Manera}, C.~{Maraston}, D.~{Margala}, C.~K. {McBride},
  J.~{Miralda-Escud{\'e}}, A.~D. {Myers}, R.~C. {Nichol}, P.~{Noterdaeme},
  S.~E. {Nuza}, M.~D. {Olmstead}, D.~{Oravetz}, I.~{P{\^a}ris},
  N.~{Padmanabhan}, N.~{Palanque-Delabrouille}, K.~{Pan},
  M.~{Pellejero-Ibanez}, W.~J. {Percival}, P.~{Petitjean}, M.~M. {Pieri},
  F.~{Prada}, B.~{Reid}, J.~{Rich}, N.~A. {Roe}, A.~J. {Ross}, N.~P. {Ross},
  G.~{Rossi}, J.~A. {Rubi{\~n}o-Mart{\'{\i}}n}, A.~G. {S{\'a}nchez},
  L.~{Samushia}, R.~T. {G{\'e}nova-Santos}, C.~G. {Sc{\'o}ccola}, D.~J.
  {Schlegel}, D.~P. {Schneider}, H.-J. {Seo}, E.~{Sheldon}, A.~{Simmons}, R.~A.
  {Skibba}, A.~{Slosar}, M.~A. {Strauss}, D.~{Thomas}, J.~L. {Tinker},
  R.~{Tojeiro}, J.~A. {Vazquez}, M.~{Viel}, D.~A. {Wake}, B.~A. {Weaver}, D.~H.
  {Weinberg}, W.~M. {Wood-Vasey}, C.~{Y{\`e}che}, I.~{Zehavi}, G.-B. {Zhao},
  and {BOSS Collaboration}.
\newblock {Cosmological implications of baryon acoustic oscillation
  measurements}.
\newblock {\em \prd}, 92(12):123516, December 2015.

\bibitem{2016A&A...594A..13P}
{Planck Collaboration}, P.~A.~R. {Ade}, N.~{Aghanim}, M.~{Arnaud},
  M.~{Ashdown}, J.~{Aumont}, C.~{Baccigalupi}, A.~J. {Banday}, R.~B.
  {Barreiro}, J.~G. {Bartlett}, and et~al.
\newblock {Planck 2015 results. XIII. Cosmological parameters}.
\newblock {\em \aap}, 594:A13, September 2016.

\bibitem{2014MNRAS.443L...6C}
T.~{Castro} and M.~{Quartin}.
\newblock {First measurement of {$\sigma$}$_{8}$ using supernova magnitudes
  only}.
\newblock {\em \mnras}, 443:L6--L10, September 2014.

\bibitem{2006PhRvD..74f3515D}
S.~{Dodelson} and A.~{Vallinotto}.
\newblock {Learning from the scatter in type Ia supernovae}.
\newblock {\em \prd}, 74(6):063515, September 2006.

\bibitem{2015MNRAS.449.2845A}
L.~{Amendola}, T.~{Castro}, V.~{Marra}, and M.~{Quartin}.
\newblock {Constraining the growth of perturbations with lensing of
  supernovae}.
\newblock {\em \mnras}, 449:2845--2852, May 2015.

\bibitem{2017MNRAS.465.2862S}
D.~{Scovacricchi}, R.~C. {Nichol}, E.~{Macaulay}, and D.~{Bacon}.
\newblock {Measuring weak lensing correlations of Type Ia supernovae}.
\newblock {\em \mnras}, 465:2862--2872, March 2017.

\bibitem{2006IJMPD..15.1753C}
E.~J. {Copeland}, M.~{Sami}, and S.~{Tsujikawa}.
\newblock {Dynamics of Dark Energy}.
\newblock {\em International Journal of Modern Physics D}, 15:1753--1935, 2006.

\bibitem{2013CQGra..30u4003T}
S.~{Tsujikawa}.
\newblock {Quintessence: a review}.
\newblock {\em Classical and Quantum Gravity}, 30(21):214003, November 2013.

\bibitem{2016ARNPS..66...95J}
A.~{Joyce}, L.~{Lombriser}, and F.~{Schmidt}.
\newblock {Dark Energy Versus Modified Gravity}.
\newblock {\em Annual Review of Nuclear and Particle Science}, 66:95--122,
  October 2016.

\bibitem{1964SvA.....8...13Z}
Y.~B. {Zel'dovich}.
\newblock {Observations in a Universe Homogeneous in the Mean}.
\newblock {\em \emph{Sov. Astron. Lett}}, 8:13, August 1964.

\bibitem{1969ApJ...155...89K}
R.~{Kantowski}.
\newblock {Corrections in the Luminosity-Redshift Relations of the Homogeneous
  Fried-Mann Models}.
\newblock {\em \apj}, 155:89, January 1969.

\bibitem{1973ApJ...180L..31D}
C.~C. {Dyer} and R.~C. {Roeder}.
\newblock {Distance-Redshift Relations for Universes with Some Intergalactic
  Medium}.
\newblock {\em \apjl}, 180:L31, February 1973.

\bibitem{2016PhRvL.116f1102A}
B.~P. {Abbott}, R.~{Abbott}, T.~D. {Abbott}, M.~R. {Abernathy}, F.~{Acernese},
  K.~{Ackley}, C.~{Adams}, T.~{Adams}, P.~{Addesso}, R.~X. {Adhikari}, and
  et~al.
\newblock {Observation of Gravitational Waves from a Binary Black Hole Merger}.
\newblock {\em Physical Review Letters}, 116(6):061102, February 2016.

\bibitem{2016PhRvL.116x1103A}
B.~P. {Abbott}, R.~{Abbott}, T.~D. {Abbott}, M.~R. {Abernathy}, F.~{Acernese},
  K.~{Ackley}, C.~{Adams}, T.~{Adams}, P.~{Addesso}, R.~X. {Adhikari}, and
  et~al.
\newblock {GW151226: Observation of Gravitational Waves from a 22-Solar-Mass
  Binary Black Hole Coalescence}.
\newblock {\em Physical Review Letters}, 116(24):241103, June 2016.

\bibitem{2017PhRvL.118v1101A}
B.~P. {Abbott}, R.~{Abbott}, T.~D. {Abbott}, F.~{Acernese}, K.~{Ackley},
  C.~{Adams}, T.~{Adams}, P.~{Addesso}, R.~X. {Adhikari}, V.~B. {Adya}, and
  et~al.
\newblock {GW170104: Observation of a 50-Solar-Mass Binary Black Hole
  Coalescence at Redshift 0.2}.
\newblock {\em Physical Review Letters}, 118(22):221101, June 2017.

\bibitem{2016PhRvD..94h3504C}
B.~{Carr}, F.~{K{\"u}hnel}, and M.~{Sandstad}.
\newblock {Primordial black holes as dark matter}.
\newblock {\em \prd}, 94(8):083504, October 2016.

\bibitem{2001ApJ...559...53M}
E.~{M{\"o}rtsell}, A.~{Goobar}, and L.~{Bergstr{\"o}m}.
\newblock {Determining the Fraction of Compact Objects in the Universe Using
  Supernova Observations}.
\newblock {\em \apj}, 559:53--58, September 2001.

\bibitem{2017Sci...356..291G}
A.~{Goobar}, R.~{Amanullah}, S.~R. {Kulkarni}, P.~E. {Nugent}, J.~{Johansson},
  C.~{Steidel}, D.~{Law}, E.~{M{\"o}rtsell}, R.~{Quimby}, N.~{Blagorodnova},
  A.~{Brandeker}, Y.~{Cao}, A.~{Cooray}, R.~{Ferretti}, C.~{Fremling},
  L.~{Hangard}, M.~{Kasliwal}, T.~{Kupfer}, R.~{Lunnan}, F.~{Masci}, A.~A.
  {Miller}, H.~{Nayyeri}, J.~D. {Neill}, E.~O. {Ofek}, S.~{Papadogiannakis},
  T.~{Petrushevska}, V.~{Ravi}, J.~{Sollerman}, M.~{Sullivan}, F.~{Taddia},
  R.~{Walters}, D.~{Wilson}, L.~{Yan}, and O.~{Yaron}.
\newblock {iPTF16geu: A multiply imaged, gravitationally lensed type Ia
  supernova}.
\newblock {\em Science}, 356:291--295, April 2017.

\bibitem{2014A&A...564A.125B}
J.~{Buchner}, A.~{Georgakakis}, K.~{Nandra}, L.~{Hsu}, C.~{Rangel},
  M.~{Brightman}, A.~{Merloni}, M.~{Salvato}, J.~{Donley}, and D.~{Kocevski}.
\newblock {X-ray spectral modelling of the AGN obscuring region in the CDFS:
  Bayesian model selection and catalogue}.
\newblock {\em \aap}, 564:A125, April 2014.

\bibitem{2006astro.ph..9591A}
A.~{Albrecht}, G.~{Bernstein}, R.~{Cahn}, W.~L. {Freedman}, J.~{Hewitt},
  W.~{Hu}, J.~{Huth}, M.~{Kamionkowski}, E.~W. {Kolb}, L.~{Knox}, J.~C.
  {Mather}, S.~{Staggs}, and N.~B. {Suntzeff}.
\newblock {Report of the Dark Energy Task Force}.
\newblock {\em ArXiv Astrophysics e-prints}, September 2006.

\bibitem{2009ApJS..180..330K}
E.~{Komatsu}, J.~{Dunkley}, M.~R. {Nolta}, C.~L. {Bennett}, B.~{Gold},
  G.~{Hinshaw}, N.~{Jarosik}, D.~{Larson}, M.~{Limon}, L.~{Page}, D.~N.
  {Spergel}, M.~{Halpern}, R.~S. {Hill}, A.~{Kogut}, S.~S. {Meyer}, G.~S.
  {Tucker}, J.~L. {Weiland}, E.~{Wollack}, and E.~L. {Wright}.
\newblock {Five-Year Wilkinson Microwave Anisotropy Probe Observations:
  Cosmological Interpretation}.
\newblock {\em \apjs}, 180:330--376, February 2009.

\bibitem{corner}
Daniel Foreman-Mackey.
\newblock corner.py: Scatterplot matrices in python.
\newblock {\em The Journal of Open Source Software}, 24, 2016.

\bibitem{2011MNRAS.416.3017B}
F.~{Beutler}, C.~{Blake}, M.~{Colless}, D.~H. {Jones}, L.~{Staveley-Smith},
  L.~{Campbell}, Q.~{Parker}, W.~{Saunders}, and F.~{Watson}.
\newblock {The 6dF Galaxy Survey: baryon acoustic oscillations and the local
  Hubble constant}.
\newblock {\em \mnras}, 416:3017--3032, October 2011.

\bibitem{2012MNRAS.427.2132P}
N.~{Padmanabhan}, X.~{Xu}, D.~J. {Eisenstein}, R.~{Scalzo}, A.~J. {Cuesta},
  K.~T. {Mehta}, and E.~{Kazin}.
\newblock {A 2 per cent distance to z = 0.35 by reconstructing baryon acoustic
  oscillations - I. Methods and application to the Sloan Digital Sky Survey}.
\newblock {\em \mnras}, 427:2132--2145, December 2012.

\bibitem{2012MNRAS.427.3435A}
L.~{Anderson}, E.~{Aubourg}, S.~{Bailey}, D.~{Bizyaev}, M.~{Blanton}, A.~S.
  {Bolton}, J.~{Brinkmann}, J.~R. {Brownstein}, A.~{Burden}, A.~J. {Cuesta},
  L.~A.~N. {da Costa}, K.~S. {Dawson}, R.~{de Putter}, D.~J. {Eisenstein},
  J.~E. {Gunn}, H.~{Guo}, J.-C. {Hamilton}, P.~{Harding}, S.~{Ho},
  K.~{Honscheid}, E.~{Kazin}, D.~{Kirkby}, J.-P. {Kneib}, A.~{Labatie},
  C.~{Loomis}, R.~H. {Lupton}, E.~{Malanushenko}, V.~{Malanushenko},
  R.~{Mandelbaum}, M.~{Manera}, C.~{Maraston}, C.~K. {McBride}, K.~T. {Mehta},
  O.~{Mena}, F.~{Montesano}, D.~{Muna}, R.~C. {Nichol}, S.~E. {Nuza}, M.~D.
  {Olmstead}, D.~{Oravetz}, N.~{Padmanabhan}, N.~{Palanque-Delabrouille},
  K.~{Pan}, J.~{Parejko}, I.~{P{\^a}ris}, W.~J. {Percival}, P.~{Petitjean},
  F.~{Prada}, B.~{Reid}, N.~A. {Roe}, A.~J. {Ross}, N.~P. {Ross},
  L.~{Samushia}, A.~G. {S{\'a}nchez}, D.~J. {Schlegel}, D.~P. {Schneider},
  C.~G. {Sc{\'o}ccola}, H.-J. {Seo}, E.~S. {Sheldon}, A.~{Simmons}, R.~A.
  {Skibba}, M.~A. {Strauss}, M.~E.~C. {Swanson}, D.~{Thomas}, J.~L. {Tinker},
  R.~{Tojeiro}, M.~V. {Maga{\~n}a}, L.~{Verde}, C.~{Wagner}, D.~A. {Wake},
  B.~A. {Weaver}, D.~H. {Weinberg}, M.~{White}, X.~{Xu}, C.~{Y{\`e}che},
  I.~{Zehavi}, and G.-B. {Zhao}.
\newblock {The clustering of galaxies in the SDSS-III Baryon Oscillation
  Spectroscopic Survey: baryon acoustic oscillations in the Data Release 9
  spectroscopic galaxy sample}.
\newblock {\em \mnras}, 427:3435--3467, December 2012.

\bibitem{2014MNRAS.441.3524K}
E.~A. {Kazin}, J.~{Koda}, C.~{Blake}, N.~{Padmanabhan}, S.~{Brough},
  M.~{Colless}, C.~{Contreras}, W.~{Couch}, S.~{Croom}, D.~J. {Croton}, T.~M.
  {Davis}, M.~J. {Drinkwater}, K.~{Forster}, D.~{Gilbank}, M.~{Gladders},
  K.~{Glazebrook}, B.~{Jelliffe}, R.~J. {Jurek}, I.-h. {Li}, B.~{Madore}, D.~C.
  {Martin}, K.~{Pimbblet}, G.~B. {Poole}, M.~{Pracy}, R.~{Sharp},
  E.~{Wisnioski}, D.~{Woods}, T.~K. {Wyder}, and H.~K.~C. {Yee}.
\newblock {The WiggleZ Dark Energy Survey: improved distance measurements to z
  = 1 with reconstruction of the baryonic acoustic feature}.
\newblock {\em \mnras}, 441:3524--3542, July 2014.

\bibitem{1965SvA.....8..854D}
V.~M. {Dashevskii} and Y.~B. {Zel'dovich}.
\newblock {Propagation of Light in a Nonhomogeneous Nonflat Universe II}.
\newblock {\em \emph{Sov. Astron. Lett}}, 8:854, June 1965.

\bibitem{1966SvA.....9..671D}
V.~M. {Dashevskii} and V.~I. {Slysh}.
\newblock {On the Propagation of Light in a Nonhomogeneous Universe}.
\newblock {\em \emph{Sov. Astron. Lett}}, 9:671, February 1966.

\bibitem{1966RSPSA.294..195B}
B.~{Bertotti}.
\newblock {The Luminosity of Distant Galaxies}.
\newblock {\em Proceedings of the Royal Society of London Series A},
  294:195--207, September 1966.

\bibitem{1967ApJ...150..737G}
J.~E. {Gunn}.
\newblock {On the Propagation of Light in Inhomogeneous Cosmologies. I. Mean
  Effects}.
\newblock {\em \apj}, 150:737, December 1967.

\bibitem{1992grle.book.....S}
P.~{Schneider}, J.~{Ehlers}, and E.~E. {Falco}.
\newblock {\em {Gravitational Lenses}}.
\newblock 1992.

\bibitem{2013PhRvD..87f3527B}
N.~{Bret{\'o}n} and A.~{Montiel}.
\newblock {Observational constraints from supernovae Ia and gamma-ray bursts on
  a clumpy universe}.
\newblock {\em \prd}, 87(6):063527, March 2013.

\bibitem{2015MNRAS.451.2097H}
P.~{Helbig}.
\newblock {The m-z relation for Type Ia supernovae, locally inhomogeneous
  cosmological models, and the nature of dark matter}.
\newblock {\em \mnras}, 451:2097--2107, August 2015.

\bibitem{2001IJMPD..10..213C}
M.~{Chevallier} and D.~{Polarski}.
\newblock {Accelerating Universes with Scaling Dark Matter}.
\newblock {\em International Journal of Modern Physics D}, 10:213--223, 2001.

\bibitem{2003PhRvL..90i1301L}
E.~V. {Linder}.
\newblock {Exploring the Expansion History of the Universe}.
\newblock {\em Physical Review Letters}, 90(9):091301, March 2003.

\bibitem{1933PMag...15..761E}
I.~M.~H. {Etherington}.
\newblock {On the Definition of Distance in General Relativity.}
\newblock {\em Philosophical Magazine}, 15, 1933.

\bibitem{2004PhRvD..69j1305B}
B.~A. {Bassett} and M.~{Kunz}.
\newblock {Cosmic distance-duality as a probe of exotic physics and
  acceleration}.
\newblock {\em \prd}, 69(10):101305, May 2004.

\bibitem{2011JCAP...05..023N}
R.~{Nair}, S.~{Jhingan}, and D.~{Jain}.
\newblock {Observational cosmology and the cosmic distance duality relation}.
\newblock {\em \jcap}, 5:023, May 2011.

\bibitem{2016arXiv161109426H}
R.~F.~L. {Holanda}, V.~C. {Busti}, F.~S. {Lima}, and J.~S. {Alcaniz}.
\newblock {Probing the distance-duality relation with high-$z$ data}.
\newblock {\em ArXiv e-prints}, November 2016.

\bibitem{1976ApJ...208L...1W}
S.~{Weinberg}.
\newblock {Apparent luminosities in a locally inhomogeneous universe}.
\newblock {\em \apjl}, 208:L1--L3, August 1976.

\bibitem{2012MNRAS.426.1121C}
C.~{Clarkson}, G.~F.~R. {Ellis}, A.~{Faltenbacher}, R.~{Maartens}, O.~{Umeh},
  and J.-P. {Uzan}.
\newblock {(Mis)interpreting supernovae observations in a lumpy universe}.
\newblock {\em \mnras}, 426:1121--1136, October 2012.

\bibitem{2017arXiv170906022M}
F.~{Montanari} and S.~{Rasanen}.
\newblock {Backreaction and FRW consistency conditions}.
\newblock {\em ArXiv e-prints}, September 2017.

\bibitem{1999A&A...351L..10S}
U.~{Seljak} and D.~E. {Holz}.
\newblock {Limits on the density of compact objects from high redshift
  supernovae}.
\newblock {\em \aap}, 351:L10--L14, November 1999.

\bibitem{2002A&A...382..787M}
E.~{M{\"o}rtsell}.
\newblock {The Dyer-Roeder distance-redshift relation in inhomogeneous
  universes}.
\newblock {\em \aap}, 382:787--791, February 2002.

\bibitem{2007PhRvL..98g1302M}
R.~B. {Metcalf} and J.~{Silk}.
\newblock {New Constraints on Macroscopic Compact Objects as Dark Matter
  Candidates from Gravitational Lensing of Type Ia Supernovae}.
\newblock {\em Physical Review Letters}, 98(7):071302, February 2007.

\bibitem{2002A&A...392..757G}
A.~{Goobar}, E.~{M{\"o}rtsell}, R.~{Amanullah}, M.~{Goliath},
  L.~{Bergstr{\"o}m}, and T.~{Dahl{\'e}n}.
\newblock {SNOC: A Monte-Carlo simulation package for high-z supernova
  observations}.
\newblock {\em \aap}, 392:757--771, September 2002.

\bibitem{2011ApJS..192....1C}
A.~{Conley}, J.~{Guy}, M.~{Sullivan}, N.~{Regnault}, P.~{Astier}, C.~{Balland},
  S.~{Basa}, R.~G. {Carlberg}, D.~{Fouchez}, D.~{Hardin}, I.~M. {Hook}, D.~A.
  {Howell}, R.~{Pain}, N.~{Palanque-Delabrouille}, K.~M. {Perrett}, C.~J.
  {Pritchet}, J.~{Rich}, V.~{Ruhlmann-Kleider}, D.~{Balam}, S.~{Baumont}, R.~S.
  {Ellis}, S.~{Fabbro}, H.~K. {Fakhouri}, N.~{Fourmanoit},
  S.~{Gonz{\'a}lez-Gait{\'a}n}, M.~L. {Graham}, M.~J. {Hudson}, E.~{Hsiao},
  T.~{Kronborg}, C.~{Lidman}, A.~M. {Mourao}, J.~D. {Neill}, S.~{Perlmutter},
  P.~{Ripoche}, N.~{Suzuki}, and E.~S. {Walker}.
\newblock {Supernova Constraints and Systematic Uncertainties from the First
  Three Years of the Supernova Legacy Survey}.
\newblock {\em \apjs}, 192:1, January 2011.

\bibitem{2011ApJ...737..102S}
M.~{Sullivan}, J.~{Guy}, A.~{Conley}, N.~{Regnault}, P.~{Astier}, C.~{Balland},
  S.~{Basa}, R.~G. {Carlberg}, D.~{Fouchez}, D.~{Hardin}, I.~M. {Hook}, D.~A.
  {Howell}, R.~{Pain}, N.~{Palanque-Delabrouille}, K.~M. {Perrett}, C.~J.
  {Pritchet}, J.~{Rich}, V.~{Ruhlmann-Kleider}, D.~{Balam}, S.~{Baumont}, R.~S.
  {Ellis}, S.~{Fabbro}, H.~K. {Fakhouri}, N.~{Fourmanoit},
  S.~{Gonz{\'a}lez-Gait{\'a}n}, M.~L. {Graham}, M.~J. {Hudson}, E.~{Hsiao},
  T.~{Kronborg}, C.~{Lidman}, A.~M. {Mourao}, J.~D. {Neill}, S.~{Perlmutter},
  P.~{Ripoche}, N.~{Suzuki}, and E.~S. {Walker}.
\newblock {SNLS3: Constraints on Dark Energy Combining the Supernova Legacy
  Survey Three-year Data with Other Probes}.
\newblock {\em \apj}, 737:102, August 2011.

\bibitem{2017arXiv171202240Z}
Miguel {Zumalacarregui} and Uros {Seljak}.
\newblock {No LIGO MACHO: Primordial Black Holes, Dark Matter and Gravitational
  Lensing of Type Ia Supernovae}.
\newblock {\em ArXiv e-prints}, page arXiv:1712.02240, December 2017.

\bibitem{2018PDU....20...95G}
Juan {Garc{\'\i}a-Bellido}, S{\'e}bastien {Clesse}, and Pierre {Fleury}.
\newblock {Primordial black holes survive SN lensing constraints}.
\newblock {\em Physics of the Dark Universe}, 20:95--100, June 2018.

\bibitem{2014A&A...572A..80A}
P.~{Astier}, C.~{Balland}, M.~{Brescia}, E.~{Cappellaro}, R.~G. {Carlberg},
  S.~{Cavuoti}, M.~{Della Valle}, E.~{Gangler}, A.~{Goobar}, J.~{Guy},
  D.~{Hardin}, I.~M. {Hook}, R.~{Kessler}, A.~{Kim}, E.~{Linder}, G.~{Longo},
  K.~{Maguire}, F.~{Mannucci}, S.~{Mattila}, R.~{Nichol}, R.~{Pain},
  N.~{Regnault}, S.~{Spiro}, M.~{Sullivan}, C.~{Tao}, M.~{Turatto}, X.~F.
  {Wang}, and W.~M. {Wood-Vasey}.
\newblock {Extending the supernova Hubble diagram to z \~{} 1.5 with the Euclid
  space mission}.
\newblock {\em \aap}, 572:A80, December 2014.

\bibitem{2013arXiv1305.5422S}
D.~{Spergel}, N.~{Gehrels}, J.~{Breckinridge}, M.~{Donahue}, A.~{Dressler},
  B.~S. {Gaudi}, T.~{Greene}, O.~{Guyon}, C.~{Hirata}, J.~{Kalirai}, N.~J.
  {Kasdin}, W.~{Moos}, S.~{Perlmutter}, M.~{Postman}, B.~{Rauscher},
  J.~{Rhodes}, Y.~{Wang}, D.~{Weinberg}, J.~{Centrella}, W.~{Traub},
  C.~{Baltay}, J.~{Colbert}, D.~{Bennett}, A.~{Kiessling}, B.~{Macintosh},
  J.~{Merten}, M.~{Mortonson}, M.~{Penny}, E.~{Rozo}, D.~{Savransky},
  K.~{Stapelfeldt}, Y.~{Zu}, C.~{Baker}, E.~{Cheng}, D.~{Content}, J.~{Dooley},
  M.~{Foote}, R.~{Goullioud}, K.~{Grady}, C.~{Jackson}, J.~{Kruk}, M.~{Levine},
  M.~{Melton}, C.~{Peddie}, J.~{Ruffa}, and S.~{Shaklan}.
\newblock {Wide-Field InfraRed Survey Telescope-Astrophysics Focused Telescope
  Assets WFIRST-AFTA Final Report}.
\newblock {\em ArXiv e-prints}, May 2013.

\bibitem{2012JCAP...11..045B}
I.~{Ben-Dayan}, G.~{Marozzi}, F.~{Nugier}, and G.~{Veneziano}.
\newblock {The second-order luminosity-redshift relation in a generic
  inhomogeneous cosmology}.
\newblock {\em \jcap}, 11:045, November 2012.

\bibitem{2013PhRvL.110b1301B}
I.~{Ben-Dayan}, M.~{Gasperini}, G.~{Marozzi}, F.~{Nugier}, and G.~{Veneziano}.
\newblock {Do Stochastic Inhomogeneities Affect Dark-Energy Precision
  Measurements?}
\newblock {\em Physical Review Letters}, 110(2):021301, January 2013.

\bibitem{2017JCAP...03..062F}
P.~{Fleury}, C.~{Clarkson}, and R.~{Maartens}.
\newblock {How does the cosmic large-scale structure bias the Hubble diagram?}
\newblock {\em \jcap}, 3:062, March 2017.

\bibitem{2013PhRvD..87l3526F}
P.~{Fleury}, H.~{Dupuy}, and J.-P. {Uzan}.
\newblock {Interpretation of the Hubble diagram in a nonhomogeneous universe}.
\newblock {\em \prd}, 87(12):123526, June 2013.

\bibitem{2014PhRvD..90l3536P}
A.~{Peel}, M.~A. {Troxel}, and M.~{Ishak}.
\newblock {Effect of inhomogeneities on high precision measurements of
  cosmological distances}.
\newblock {\em \prd}, 90(12):123536, December 2014.

\bibitem{2014JCAP...06..054F}
P.~{Fleury}.
\newblock {Swiss-cheese models and the Dyer-Roeder approximation}.
\newblock {\em \jcap}, 6:054, June 2014.

\bibitem{2009JPhCS.189a2011E}
G.~F.~R. {Ellis}.
\newblock {Dark energy and inhomogeneity}.
\newblock In {\em Journal of Physics Conference Series}, volume 189 of {\em
  Journal of Physics Conference Series}, page 012011, October 2009.

\bibitem{2013JCAP...06..007Y}
X.~{Yang}, H.-R. {Yu}, and T.-J. {Zhang}.
\newblock {Constraining smoothness parameter and the DD relation of Dyer-Roeder
  equation with supernovae}.
\newblock {\em \jcap}, 6:007, June 2013.

\bibitem{2015PhRvD..91h3010L}
Zhengxiang {Li}, Xuheng {Ding}, and Zong-Hong {Zhu}.
\newblock {Unbiased constraints on the clumpiness of the Universe from standard
  candles}.
\newblock {\em \prd}, 91:083010, April 2015.

\end{thebibliography}

\end{document}